\documentclass[prb,amsmath,amssymb,twocolumn,superscriptaddress]{revtex4-2}
\usepackage[english]{babel}
\usepackage{graphicx}
\usepackage{epsfig}
\usepackage{dcolumn}
\usepackage{bm}
\usepackage{rotating}
\usepackage{multirow}
\usepackage{xspace}
\usepackage{dsfont}
\usepackage[colorlinks=true, allcolors=blue]{hyperref}
\usepackage[table]{xcolor}
\usepackage{float}
\usepackage{braket}
\usepackage{soul}
\usepackage{graphicx} 
\def\bk{{\bf k}}

\def\bq{{\bf q}}
\def\bQ{{\bf Q}}

\def\enk{{\varepsilon}}
\def\calP{{\mathcal{P}}}

\begin{document}
\title{Chiroptical Valleytronic in SnSe}
	\author{Yiming Pan}
    \email{pan@physik.uni-kiel.de}
	\affiliation
	{Institute of Theoretical Physics and Astrophysics, Kiel University, 24118 Kiel, Germany} 
    
        \author{Sotirios Fragkos}  
	\affiliation
	{Universit\'e de Bordeaux - CNRS - CEA, CELIA, UMR5107, F33405 Talence, France}  

        \author{Umberto Dellasette}  
	\affiliation
	{Universit\'e de Bordeaux - CNRS - CEA, CELIA, UMR5107, F33405 Talence, France}  

            \author{Quentin Courtade}  
	\affiliation
	{Universit\'e de Bordeaux - CNRS - CEA, CELIA, UMR5107, F33405 Talence, France} 

         \author{Dominique Descamps}  
	\affiliation
	{Universit\'e de Bordeaux - CNRS - CEA, CELIA, UMR5107, F33405 Talence, France} 

            \author{Stéphane Petit}  
	\affiliation
	{Universit\'e de Bordeaux - CNRS - CEA, CELIA, UMR5107, F33405 Talence, France} 

    	\author{Fabio Caruso}
	\affiliation
	{Institute of Theoretical Physics and Astrophysics, Kiel University, 24118 Kiel, Germany} 
    \affiliation
	{Institut of Physics, University of Rostock, 18059 Rostock, Germany}
    
    \author{Samuel Beaulieu}  
    \email{samuel.beaulieu@u-bordeaux.fr}
	\affiliation
	{Universit\'e de Bordeaux - CNRS - CEA, CELIA, UMR5107, F33405 Talence, France}
    
\begin{abstract}
Quantum geometric properties of Bloch electrons play a central role in governing light–matter interactions and transport phenomena in solids. Here, we report quantum-geometric chiroptical selection rules that generate a helicity-dependent momentum-space population imbalance in the group-IV monochalcogenide SnSe. By combining time- and angle-resolved extreme-ultraviolet photoemission spectroscopy with \textit{ab initio} real-time simulations of photoexcited carrier dynamics, we identify valley-selective excitation at opposite crystal momenta arising from the breaking of time-reversal symmetry under circularly polarized illumination. We theoretically demonstrate that the magnitude of valley circular dichroism is strongly constrained by crystal symmetry through its influence on the quantum geometric tensor, the key quantity underlying valley optical selection rules. The resulting chiroptically induced valley polarization decays on sub-picosecond timescales through intervalley scattering mediated by electron–phonon coupling. Furthermore, we show theoretically that the population imbalance between valleys at opposite momenta gives rise to a helicity-reversible transient photocurrent. These findings provide microscopic insight into how quantum geometry governs valley-selective light–matter interactions, valley polarization, and photocurrent generation in van der Waals materials exhibiting hidden and transient symmetry breaking.

\end{abstract}    
    \maketitle
The optical response of electrons in solids is governed not only by their energy spectrum but also by the symmetry and quantum geometry of their Bloch wavefunctions. In crystals, the wavepacket dynamics of Bloch electrons can acquire orbital angular momenta associated with their self-rotation and circulating motion in the crystal~\cite{Thonhauser2005, YaoXiao2008}, which is closely related to the Berry curvature (BC)~\cite{Cao2012, XiaoChangNiu_rmp}. A prominent manifestation of these quantum-geometric properties is valley-selective excitation by circularly polarized light (CPL) in two-dimensional (2D) hexagonal transition metal dichalcogenides (TMDCs, 2H-MX$_2$) and related materials~\cite{YaoXiao2008, Cao2012, XiaoLiu2012, Beaulieu2024}. In these systems, the optical transition probability depends on the valley index, enabling selective generation of nonequilibrium carrier populations in momentum space. More recently, it is proposed that the optical selection rules underlying such valley-selective excitation can be understood within the framework of the quantum geometric tensor (QGT)~\cite{provost1980}. Within this framework, interband transitions driven by linearly and circularly polarized light (LPL and CPL) are governed respectively by the quantum metric and BC, corresponding to the real and imaginary parts of the QGT~\cite{provost1980, LiLiu2026}. As a result, linear and circular dichroism can be expressed in terms of the QGT, and their extrema in momentum space are constrained by crystal symmetry~\cite{LiLiu2026}.  Consequently, measurements of the optical dichroism and its momentum dependence provide direct probes of the quantum geometry of Bloch electrons~\cite{de2017quantized, Hosur2011, Laman2005, Ji2019Spatially, ReesManna2020, KrishnaKumar2025}.

\begin{figure*}[t]
\begin{center}
\includegraphics[width=0.7\textwidth]{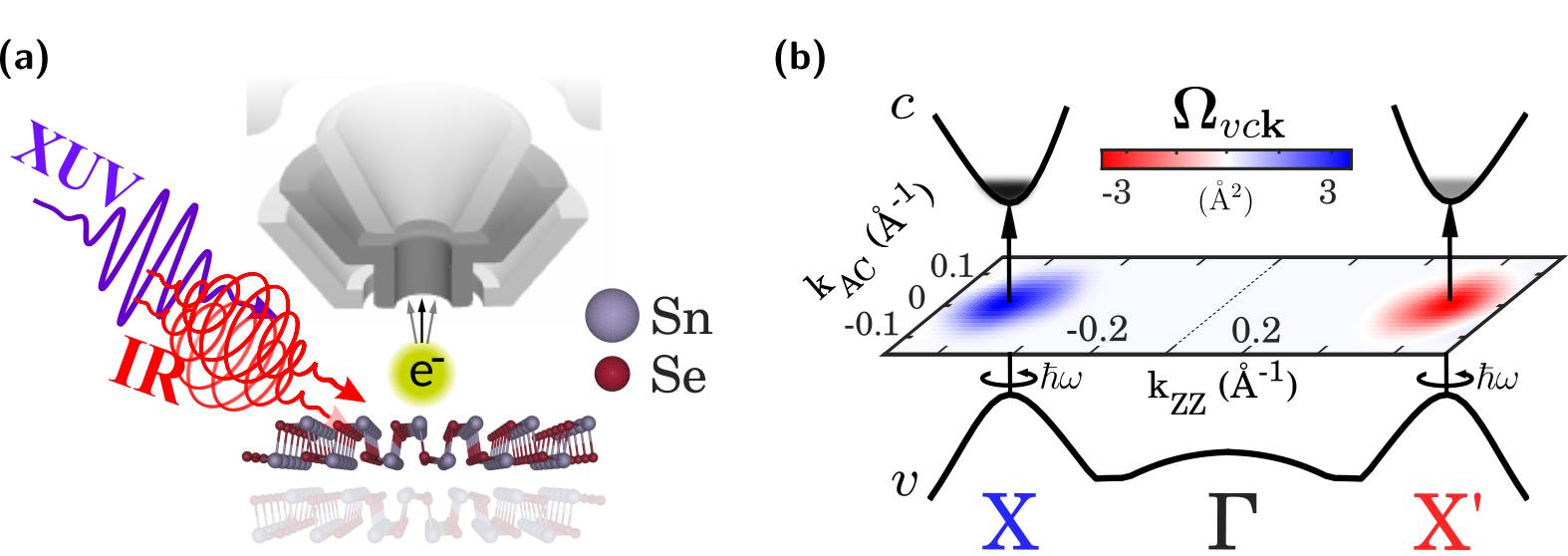}
\caption{\textbf{Experimental scheme and principle underlying chiroptical control of valley asymmetry in SnSe.} \textbf{(a)} Polarization-tunable infrared (IR) pump (1.2~eV, 135~fs, $\sim$0.48 mJ/cm$^2$) and \textit{p}-polarized XUV (21.6~eV) probe pulses are focused onto a SnSe bulk crystal inside the interaction chamber of a time-of-flight momentum microscope, where the out-of-equilibrium electronic structure and associated valley-resolved carrier populations can be measured as a function of the driving-field polarization state and pump-probe delay. \textbf{(b)} Electronic structure of SnSe along the zigzag (ZZ) direction ($k_{ZZ}$, along X-$\Gamma$-X$^{\prime}$), and calculated interband Berry curvature (BC), $\boldsymbol{\Omega}_{vc\mathbf{k}}$) around the X and X$^{\prime}$ valleys. The interband BC gives rise to optical selection rules for interband transitions driven by circularly polarized light, resulting in a helicity-dependent valley asymmetry between the X and X$^{\prime}$ valleys, highlighted by the dark and shaded carrier populations in each valley.}
\label{fig:fig1}
\end{center}
\end{figure*}

TMDC monolayers provide a well-established platform for generating and probing valley-polarized excited states using CPL~\cite{YaoXiao2008, Zeng12, Cao2012, XiaoLiu2012}. In monolayer MX$_2$, the broken inversion symmetry and time-reversal-related valleys give rise to opposite Berry curvature at the K and K$'$ points and, consequently, to nearly unity valley-selective circular dichroism~\cite{XiaoLiu2012}. However, the symmetry of pristine monolayer MX$_2$ imposes important constraints on its macroscopic response: its $D_{3h}$ point group contains a horizontal mirror plane that forbids a helicity-dependent in-plane photocurrent, despite the strong valley selectivity of the optical excitation. Thus, valley-selective excitation and photocurrent generation represent distinct physical manifestations and are not necessarily concomitant. Meanwhile, valley-selective excitation is not restricted to hexagonal materials. More generally, it can emerge whenever crystal symmetry and the quantum geometry of the electronic states produce helicity-dependent optical transition probabilities between inequivalent valleys. Group-IV monochalcogenides (MX), with their puckered crystal structures and pronounced in-plane anisotropy~\cite{Barraza-Lopez_rmp2021, WangQian2019, Panday2019, Kim2019, Hashmi2025, pan2025ultrafast}, provide an interesting platform to explore this physics beyond the conventional hexagonal-valley paradigm. 

In these systems, it has recently been predicted and demonstrated that linearly polarized light can be used to selectively drive optical transitions and generate the associated population imbalance between non-degenerate valleys located along different high-symmetry directions~\cite{Rodin2016, Lin2018, Chen18, ttNguyen2024, Hashmi2025, Hanakata2016, pan2025ultrafast}. However, exploiting quantum-geometric chiroptical selection rules to selectively address degenerate valleys with opposite Berry curvature at opposite crystal momenta remains to be demonstrated. Such a demonstration would establish group-IV monochalcogenides as a versatile platform beyond conventional valleytronic materials, which typically rely on a single valley pair. In particular, it would substantially expand the degrees of freedom available for valley manipulation, enabling both linear and circular driving fields to generate distinct forms of valley polarization, as recently discussed theoretically~\cite{Hashmi2025}.

An important question is how valley-selective excitation manifests in bulk van der Waals materials whose individual layers lack inversion symmetry, while the bulk crystal as a whole is centrosymmetric. Although inversion symmetry restores equivalence between opposite layers, each constituent layer can retain distinct momentum-space optical selection rules. Such layer-resolved, or hidden, optical responses are analogous to the hidden spin and orbital textures predicted and observed in centrosymmetric layered materials~\cite{Riley14, Bertoni2016, Razzoli17, Fanciulli2023, Beaulieu20-2, Schuler22-1}. These hidden properties can be probed using extreme-ultraviolet photoemission spectroscopy, owing to the atomic-scale inelastic mean free path of photoelectrons in this energy regime. Indeed, at photon energies around 20~eV, the inelastic mean free path of the outgoing electrons (IMFP, $\lambda$) approaches its universal minimum, with $\lambda \sim 5\textup{~\AA}$. For an interlayer spacing of $d \sim 6$--$7\textup{~\AA}$ in group-IV monochalcogenides and TMDCs, respectively, the photoemission intensity originating from the second layer is attenuated by a factor of $\exp(-d/\lambda) \approx 0.25$--$0.3$ relative to that from the topmost layer. Thus, following the established approach used to probe hidden spin and orbital textures, as well as hidden valley selection rules~\cite{Bertoni2016} in bulk TMDCs, time- and angle-resolved extreme-ultraviolet photoemission spectroscopy could enable the detection of hidden optical selection rules and the associated population imbalance in the topmost layer of bulk group-IV monochalcogenides.

Here, we combine time- and angle-resolved extreme-ultraviolet photoemission spectroscopy (tr-ARPES)~\cite{Boschini2024} with real-time simulations based on the time-dependent Boltzmann equation (TDBE)~\cite{Caruso21,PanCaruso2023,CarusoNovko2022,pan2025ultrafast} to directly investigate valley-selective excitation and the subsequent ultrafast evolution of photoexcited carriers in the group-IV monochalcogenide SnSe. The key concepts underlying our approach are illustrated in Fig.~\ref{fig:fig1}. An infrared CPL pump pulse with tunable helicity excites carriers across the band gap, while extreme-ultraviolet photoemission at variable pump-probe delays provides direct access to their energy- and momentum-resolved distribution, as illustrated in Fig.~\ref{fig:fig1}(a). Because the optical transition probabilities are helicity-dependent and the two inequivalent valleys are related by time-reversal symmetry, CPL selectively enhances the population of one valley relative to its counterpart. As a result, the photoexcited carrier distribution acquires a pronounced momentum-space asymmetry, $f_{\bk}\neq f_{-\bk}$, whose sign reverses upon reversing the helicity of the incident light, as shown in Fig.~\ref{fig:fig1}(b). We directly resolve this valley-selective population through its momentum- and time-dependent photoemission intensity and track its relaxation on the subpicosecond timescale. The experimental observations are complemented by TDBE simulations, which identify intervalley scattering mediated by electron-phonon and electron-impurity interactions as key channels governing the subsequent decay of the valley population imbalance. Together, our results establish momentum-resolved tr-ARPES as a direct probe of hidden helicity-dependent valley excitation in a centrosymmetric layered material and provide microscopic insight into the ultrafast dynamics of valley-selective light-matter coupling in group-IV monochalcogenides.

\section*{Chiroptical valley addressability}

Our experimental apparatus~\cite{Fragkos2025-ob} features a polarization-tunable ultrafast high-repetition-rate (166 kHZ) extreme ultraviolet (XUV) beamline~\cite{Comby22}, coupled to a time-of-flight momentum microscope featuring advanced multimode lens design~\cite{tkach24, tkach24-2} (see Fig.~\ref{fig:fig1}(a)). Polarization-tunable infrared (IR) pump pulses (1.2~eV, 135~fs, $\sim$0.48~mJ/cm$^2$) and \textit{p}-polarized XUV probe pulses (21.6~eV) are focused \textit{in-situ} onto a freshly cleaved SnSe sample (\textsc{HQGraphene}) at room temperature, inside the interaction chamber of the momentum microscope end station, allowing the measurement of the nonequilibrium electronic structure and occupations. More details about the setup can be found elsewhere~\cite{Fragkos2025-ob}. It is worth emphasizing that we deliberately employed a relatively low fluence ($\sim$0.48~mJ/cm$^2$), well below the threshold required to induce a nonthermal structural phase transition toward a higher-symmetry, gapless structure~\cite{Dringoli2024, Huang2022}. Consistent with this low-fluence regime, we observe no signatures of band-structure modifications that would indicate the occurrence of such a phase transition.

\begin{figure*}[t]
\begin{center}
\includegraphics[width=0.9\textwidth]{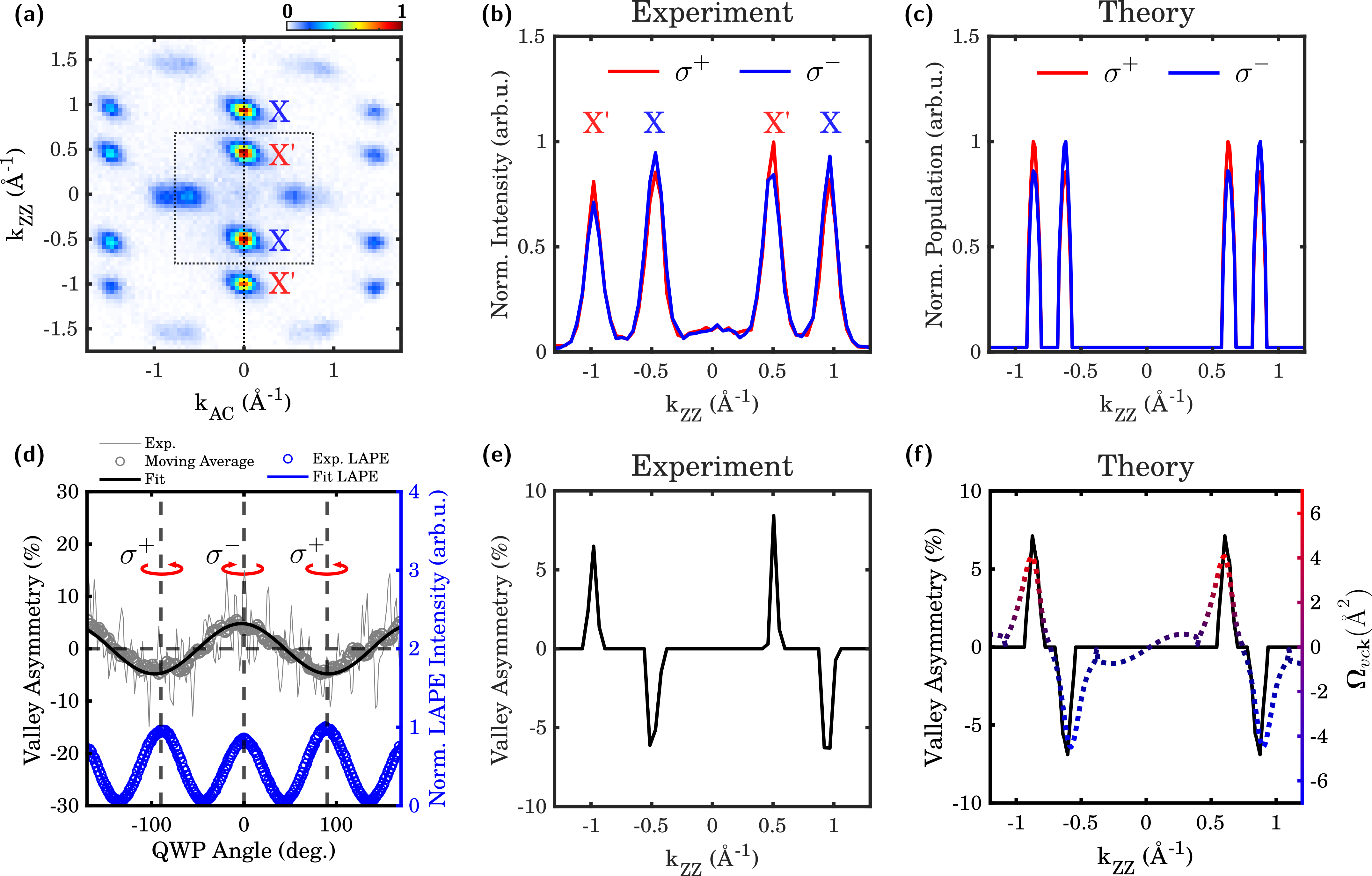}
\caption{\textbf{Helicity-dependent valley asymmetry driven by the interband Berry curvature.} 
\textbf{(a)} Constant energy contour (CEC) representing the momentum-space distribution of the conduction band (CB) population at the pump–probe temporal overlap. The thin black dashed square represents the first Brillouin zone. \textbf{(b)}-\textbf{(c)} Experimentally measured and calculated momentum distribution curves (MDCs) along the X-X$^{\prime}$-$\Gamma$-X-X$^{\prime}$ high-symmetry direction, for $\sigma^+$ and $\sigma^-$ CPL excitations, respectively. 
\textbf{(d)} Experimentally measured X/X$^{\prime}$ valley asymmetry and laser-assisted photoemission (LAPE) intensity as a function of the IR quarter-wave plate (QWP) angle, showing a reversal of the valley asymmetry upon switching the helicity of the driving pulse. \textbf{(e)}-\textbf{(f)} Experimentally measured and calculated momentum-resolved population imbalance between $\sigma^+$ and $\sigma^-$ CPL, showing opposite asymmetries between the X and X$^{\prime}$ valleys in the first and second BZs, respectively. In \textbf{(f)}, the momentum-resolved interband Berry curvature ($\boldsymbol{\Omega}_{vc\mathbf{k}}$) is superimposed (dashed curve) on the calculated valley asymmetry.}
\label{fig:fig2}
\end{center}
\end{figure*}

We first analyze valley-resolved conduction band population after CPL excitation to probe the interband optical selection rules and valley addressability. In Fig.~\ref{fig:fig2}(a), we show a constant-energy contour (CEC) representing the momentum-space distribution of the conduction band population at the pump–probe temporal overlap, integrated over all pump polarization states. Within this CEC, we observe strong populations in the X/X$^{\prime}$ valleys along $k_{ZZ}$, corresponding to the global minimum of the conduction band, and weaker populations in the Y/Y$^{\prime}$ valleys along $k_{AC}$, corresponding to higher-lying conduction band states. The thin black dashed square represents the momentum span of the first Brillouin zone. The high-energy XUV probe photons further enable the measurement of photoemission from higher-order Brillouin zones. In the following, we focus our analysis on the X–X$^{\prime}$–$\Gamma$–X–X$^{\prime}$ high-symmetry direction (around the vertical black dashed line in Fig.~\ref{fig:fig2}(a)), where we expect a helicity-dependent valley asymmetry. We then track the valley-resolved population, as a function of the IR quarter-wave plate (QWP) angle (Fig.~\ref{fig:fig2}(d)): rotating the IR QWP switches the pump polarization between $\sigma^+$ ($\pm$90°), $\sigma^-$ (0°) and s-polarized pump ($\pm$45°), revealing a reversal of the X/X$^{\prime}$ valley asymmetry (4.8$\%$ $\pm$ 1.4$\%$, 95$\%$ confidence bounds from the fit) at the helicity-swap frequency (grey and black lines in Fig.~\ref{fig:fig2}(d)). In contrast, laser-assisted photoemission (LAPE, blue dots and line) is modulated at twice this frequency, as it is maximized for each circular polarization and minimized for linear \textit{s}-polarized pump light. 

In addition, momentum distribution curves (MDCs) along this high-symmetry direction (Fig.~\ref{fig:fig2}(b)), measured under $\sigma^+$ and $\sigma^-$ circularly polarized driving pulses, clearly reveal a helicity-dependent, momentum-resolved valley asymmetry. Specifically, the conduction-band population is enhanced at the X (X$^{\prime}$) valley under $\sigma^+$ ($\sigma^-$) excitation. This behavior is further quantified in Fig.~\ref{fig:fig2}(e), which shows an opposite $\sim\pm5\%$ asymmetry between the X and X$^{\prime}$ valleys that persists across both the first and second Brillouin zones (BZs). Despite their small magnitude, these helicity-dependent valley asymmetries can be unambiguously attributed to intrinsic material properties, rather than to experimental artifacts such as an imperfect polarization state or photoemission geometry. In particular, upon translating a given valley by one reciprocal lattice vector from the first to the second BZ, the sign of the helicity-dependent asymmetry reverses, while remaining locked to the valley index (X or X$^{\prime}$), irrespective of the outgoing photoelectron emission direction. Measurements across multiple BZs therefore provide a robust means of disentangling intrinsic valley-dependent asymmetries from geometry-induced effects. In the present case, this analysis provides compelling evidence that the observed dichroism has an intrinsic origin.

To further support our experimental measurements and to understand the microscopic origin of the valley-selective optical excitation driven by CPL, we simulate the carrier dynamics in monolayer SnSe by solving the real-time TDBE. The details of the equations are provided in the Section 1 of the Supplementary Materials (SM)~\cite{Suppl}. Our real-time simulation includes both photoexcitation and carriers' relaxation processes. At the pump-probe temporal overlap, the calculated valley asymmetry is in excellent agreement with experimental measurements, as shown in Figs.~\ref{fig:fig2}(c) and~(f). To highlight its microscopic origin, the momentum-resolved interband Berry curvature ($\boldsymbol{\Omega}_{vc\mathbf{k}}$) is superimposed on the calculated valley asymmetry in Fig.~\ref{fig:fig2}(f). 

\section*{Ultrafast valley- and helicity-resolved dynamics}
\begin{figure}
\centering
\includegraphics[width=0.45\textwidth]{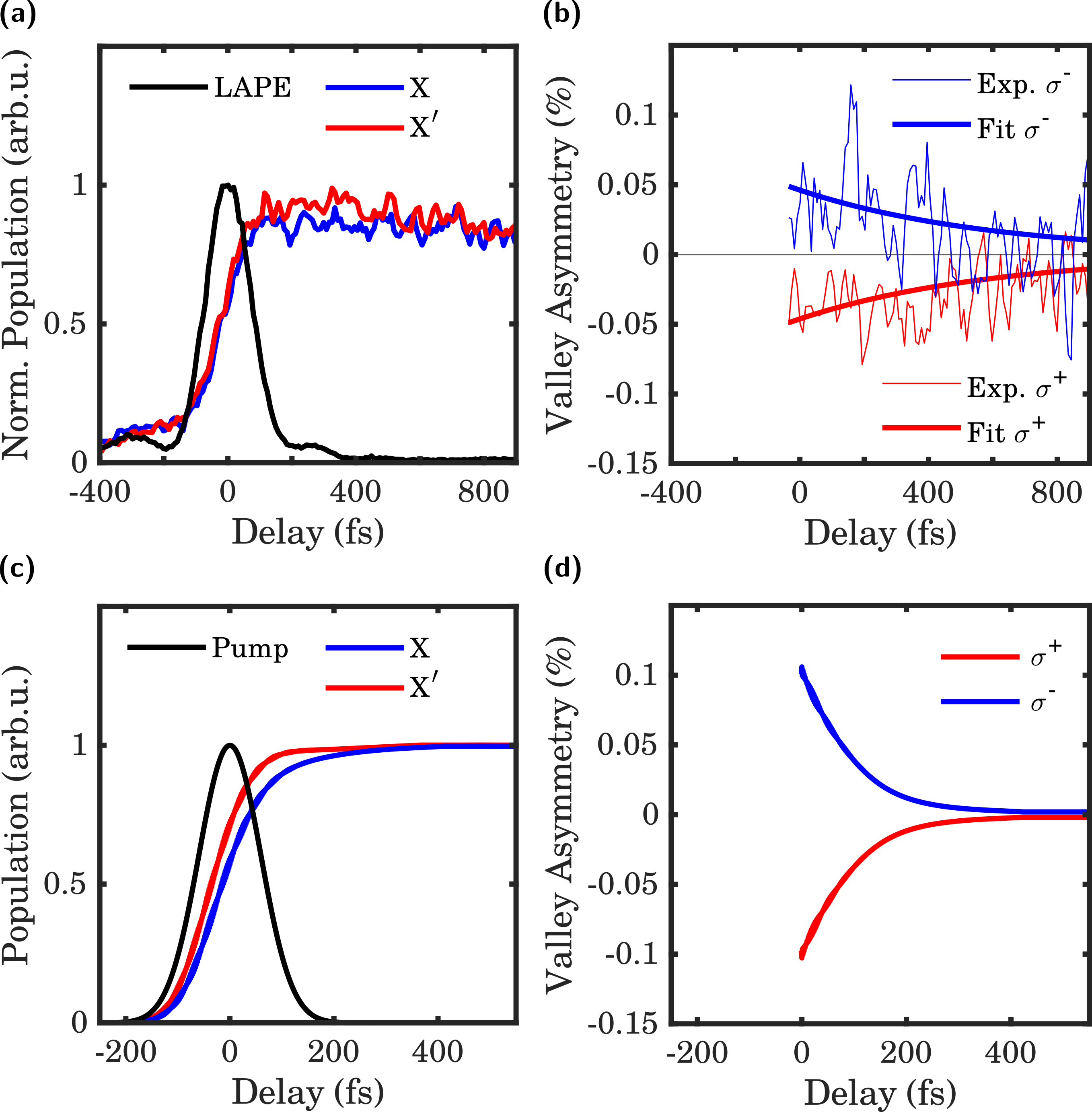}
\caption{\textbf{Ultrafast helicity- and valley-resolved dynamics.} \textbf{(a)} Experimentally measured valley-resolved (at X in blue, at X$^{\prime}$ in red and LAPE in black) carrier dynamics following photoexcitation using $\sigma^+$ CPL. \textbf{(b)} Time-resolved ultrafast X/X$^{\prime}$ valley asymmetry dynamics following photoexcitation using $\sigma^+$ (thin red line, and associated fit in thick red line) and $\sigma^-$ (thin blue line, and associated fit in thick blue line) CPL.  \textbf{(d)}-\textbf{(e)} Theoretical equivalent of \textbf{(a)}-\textbf{(b)}, \textit{i.e.} simulated with TDBE.}
\label{fig:fig3}
\end{figure}

Having established the chiroptical valley addressability in SnSe, we now investigate the ultrafast dynamics of these valley asymmetries. The initially prepared population imbalance persists over the timescale set by ultrafast intervalley scattering processes, which ultimately drive the system toward equilibrium at longer timescales. Here, using time- and valley-resolved population dynamics, we extract the ultrafast evolution of the X/X$^{\prime}$ valley asymmetry. In Fig.~\ref{fig:fig3}(a), we show the time- and valley-resolved population dynamics (in X/X$^{\prime}$ valleys in blue/red), following photoexcitation using $\sigma^+$ CPL. The valley-resolved population dynamics using the opposite pump helicity ($\sigma^-$) are shown in Fig.~S5 of Section III of the SM~\cite{Suppl}. Following impulsive interband transitions using $\sigma^+$ CPL, long-lived conduction band populations are measured at X/X$^{\prime}$ valleys. At early time delays, a slightly larger population is measured at X$^{\prime}$ valleys, with respect to X. As shown in Fig.~\ref{fig:fig3}(b), this asymmetry is (-4.6$\%$ $\pm$ 0.3$\%$, 95$\%$ confidence bounds from the fit), in good agreement with static data presented in Fig.~\ref{fig:fig2}. This asymmetry decays with a time constant of $\tau = 603 \pm 78$~fs (95$\%$ confidence interval from the exponential fit) and reverses sign upon switching the pump helicity. This behavior suggests that, on a sub-picosecond timescale, directional intervalley scattering progressively erases the optically induced valley asymmetry. Our TDBE calculations qualitatively reproduce this behavior (see Fig.~\ref{fig:fig3}(c)--(d)), with the faster decay predicted by the simulations likely resulting from the gradient approximation employed in TDBE, which is known to overestimate carrier relaxation rates~\cite{Picano2021,Girotto2025}. Furthermore, as discussed in the next section, these simulations provide access to the phonon mode mediating this intervalley scattering process, as well as to the magnitude and dynamics of the transient photocurrent associated with the resulting momentum-space population imbalance.

\section*{Generation and decay of transient photocurrent}
As discussed in pioneering time- and angle-resolved photoemission studies of metal surfaces~\cite{gudde07} and topological insulators~\cite{Soifer2019}, an asymmetric electron population at opposite crystal momenta in dispersive bands can be regarded as a spectroscopic signature of a photocurrent. In the following, we theoretically investigate how the measured chiroptical valley-selection rules and the associated helicity-dependent population imbalance between opposite crystal momenta could give rise to a directional, transient photocurrent in SnSe.

\begin{figure*}[t]
\begin{center}
\includegraphics[width=0.8\textwidth]{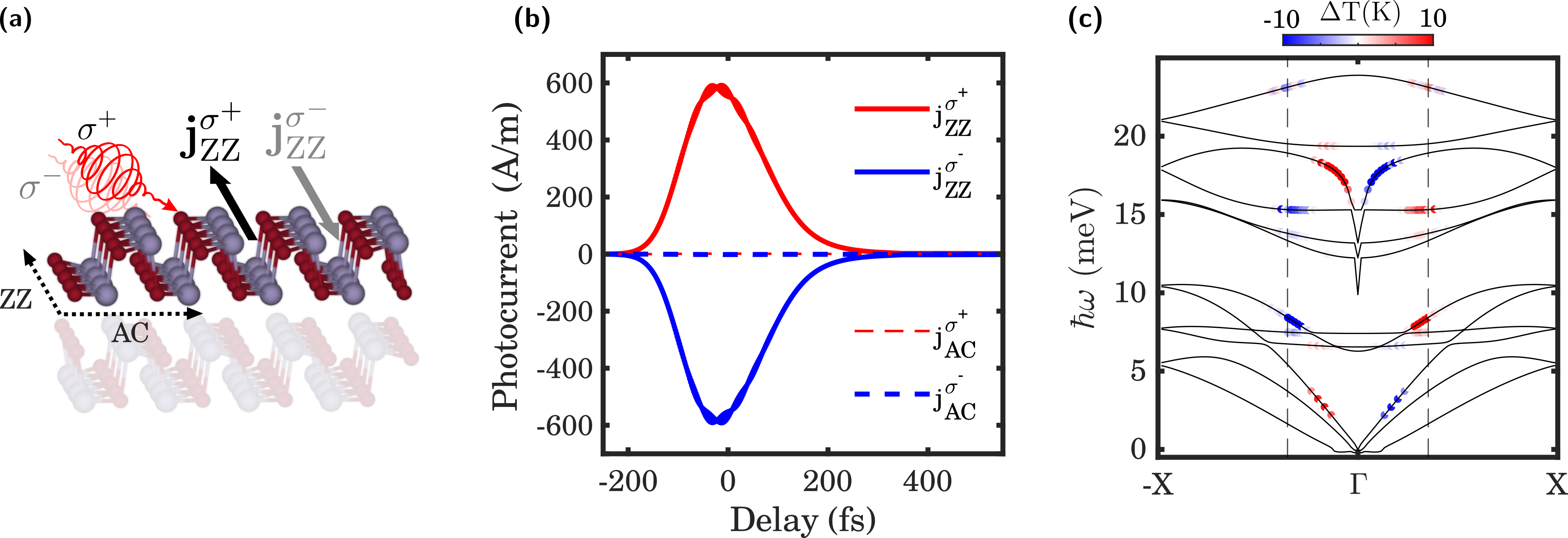}
\caption{\textbf{Generation and mode-resolved phonon-induced decay of hidden photocurrent in SnSe.} \textbf{(a)} Schematic representation of the directional and helicity-dependent photocurrent along the zigzag direction ($j_{ZZ}^{\sigma^+} = -j_{ZZ }^{\sigma^-}$) due to X/X$^{\prime}$ valley asymmetry. ZZ and AC represent the zigzag and armchair crystallographic axes. \textbf{(b)} Helicity-dependent directional transient photocurrent associated with X/X$^{\prime}$ valley asymmetry, calculated using Eq.~\eqref{eq:photocurrent}. \textbf{(c)} Mode-resolved differential phonon temperature along the X$^{\prime}$-$\Gamma$-X high-symmetry direction, obtained by subtracting the phonon temperature under photoexcitation using $\sigma^+$ and $\sigma^-$ CPL.}
\label{fig:fig4}
\end{center}
\end{figure*}

Under time-reversal symmetry, the group velocities of carriers are constrained by ${\bf v}(\bk) = -{\bf v}(-\bk)$, while with mirror-plane symmetry along the armchair direction $M_{yz}$, we have $v_x(k_x,k_y)=-v_x(-k_x,k_y)$ and $v_y(k_x,k_y)=v_y(-k_x,k_y)$. As a result, for the electronic population, the group velocities satisfy $v_x(k_x,0)=-v_x(-k_x,0)$ and $v_y(k_x,0)=0$, and the imbalanced photoexcited carriers give rise to a photocurrent flowing along the zigzag ($x$) direction, as schematically sketched in Fig.~\ref{fig:fig4}(a). 

Moreover, photocurrent can be evaluated from the band velocity ${\bf v}_{n\bk}=\hbar^{-1}\nabla_{\bk}\enk_{n\bk}$ weighted by the carrier population $f_{n\bk}(t)$,
\begin{align}  
{\bf j} =(N_{\bk}S_{2D})^{-1}\sum_{n\bk}f_{n\bk}(t) {\bf v}_{n\bk} \label{eq:photocurrent},   
\end{align}
where $N_{\bk}S_{2D}$ represents the area normal to the direction of the current. The simulations presented in Fig.~\ref{fig:fig4}(b) reveal a subpicosecond transient photocurrent along the ZZ direction (vanishing along AC), whose sign reverses with CPL helicity. Direct experimental extraction of this current would require knowledge of the conduction band dispersion, which is not possible here since only the bottom of this band is significantly populated.

In addition, momentum conservation in intervalley scattering allows us to identify the phonon modes responsible for the decay of momentum-space population asymmetry and associated photocurrent. From the band structure of monolayer SnSe (Fig. S2), the X valleys appear at $\bk_0\approx(\pm0.412\frac{2\pi}{a_x},0)$, and the population imbalance between the opposite X valleys can thus be erased by phonon emission/absorption with momentum $\bq_0=2|\bk_0|\pm {\bf G}\approx (\pm0.176\frac{2\pi}{a_x},0)$, represented by vertical dashed lines in Fig.~\ref{fig:fig4}(c). We can quantify the difference of nonthermal phonon established in the electron-phonon scattering process following $\sigma^+$ and $\sigma^-$ excitation, \textit{i.e.} $\Delta T_{\bq\nu}(t) = T_{\bq\nu} ^{\sigma+}(t) - T^{\sigma-}_{\bq\nu}(t)$, where the nonequilibrium phonon population $n_{\bq\nu}$ is converted to an effective temperature~\cite{Caruso21} by inverting the Bose-Einstein function $T_{\bq\nu} = \hbar\omega_{\bq\nu}\left[ k_B{\rm ln}(1+n_{\bq\nu}^{-1})\right]^{-1}$, as shown in Fig.~\ref{fig:fig4}(c). Dichroic phonon temperatures emerge around $\bq_0=2|\bk_0|\pm {\bf G}\approx (\pm0.176\frac{2\pi}{a_x},0)$, at $\sim$ 8~meV and 16~meV, associated with mode-specific phonon emission/absorption responsible for helicity-dependent relaxation of momentum-space population asymmetry. Time- and helicity-resolved diffuse electron scattering~\cite{Seiler21, Britt2024, Britt2023} on monolayer SnSe could provide a powerful probe of the mode-specific dichroic phonon temperatures underlying the dynamics of chiroptical valleytronics, making it a promising avenue for future investigation.
 
It is important to emphasize that the momentum-space population asymmetry and the associated transient photocurrent are nonzero in monolayer SnSe, as well as within each individual layer of bulk SnSe, but reverse sign between adjacent layers. We confirmed numerically that, due to this layer-by-layer cancellation, the CPL-induced population asymmetry and the resulting photocurrent vanish in bulk SnSe.

\section*{Quantum geometric origin of chiroptical selection rules}
In the following, we turn our attention to the detailed physical origin of the chiroptical selection rules and the associated, small but non-vanishing, valley asymmetries induced by CPL excitation in SnSe. Importantly, even at the theoretical level, the valley asymmetry between the X and X' valleys in SnSe induced by CPL is found to be substantially smaller than the corresponding asymmetry between the K and K' valleys in TMDCs under identical CPL excitation~\cite{Cao2012}.

While phonon-assisted intervalley scattering dynamically reduces the valley asymmetry, its upper bound is given by the degree of circular dichroism~\cite{YaoXiao2008,Cao2012}:
\begin{align}
  \eta_{\bk} = \frac{|\calP_{cv\bk}^{+}|^2-|\calP_{cv\bk}^{-}|^2}{|\calP_{cv\bk}^{+}|^2+|\calP_{cv\bk}^{-}|^2} , \label{eq:DCD}  
\end{align}
where the interband dipole $\calP_{cv\bk}^{\pm} =\bra{\psi_{v\bk}}\hat{p}_x\pm i\hat{p}_y\ket{\psi_{c\bk}}$ represents the transition amplitude between valence and conduction bands under $\sigma^+$ and $\sigma^-$ CPL, with $\hat{p}_{x/y}$ the $x/y$ component of the electronic momentum operator. Eq.~\eqref{eq:DCD} defines the maximum value of valley asymmetry without any scattering and is determined by the valleys' symmetry and orbital properties. Eq.~\eqref{eq:DCD} can also be recast into $\eta_{\bk} = -\frac{\Omega_{vc\bk}^{xy}}{g_{vc\bk}^{xx}+g_{vc\bk}^{yy}},$ where the interband BC, quantum metric and QGT are defined by $\Omega_{vc\bk}^{xy}=-2{\rm Im}\left[r_{vc\bk}^xr_{cv\bk}^{y}\right]$, $g_{vc\bk}^{ij}={\rm Re}\left[r_{vc\bk}^ir_{cv\bk}^j\right]$ ($i,j =x,y$) and $\bQ_{vc\bk} = {\bf g}_{vc\bk} -i\mathbf{\Omega}_{vc\bk}/2$~\cite{LiLiu2026}, and  the inequality $|\eta_{\bk}|\leq 1$ follows from the semi-definiteness of QGT, i.e. $tr(g) = g_{vc\bk}^{xx}+g_{vc\bk}^{yy}\geq |\Omega^{xy}_{vc\bk}|$~\cite{LiLiu2026,Roy2014}. Moreover, $\eta_{\bk} = 1$ is satisfied only for valleys with $n-$fold ($n\geq3$) rotational symmetry~\cite{LiLiu2026}, which is the case in TMDCs. For monolayer SnSe, only screw rotational symmetry $\bar{C}_{2y}$ with $n=2$ is present; thus, it is not surprising to compute and measure an incomplete, i.e., $<$ 100\% circular dichroism. 

To gain deeper quantitative insight into the circular dichroism of SnSe observed both in experiments and \textit{ab initio} simulations, we derive a tight-binding model for the X-valley physics. From previous DFT calculations, it is known that electronic states around the X-valley are made essentially of $p_x$ orbitals of the Sn (Se) atoms for the electronic states in the conduction (valence) bands~\cite{ttNguyen2024, Kim2019, ZhengZhang2024}. Using $p_x$ orbitals of each atomic site and the glide-reflection symmetry $\bar{M}_z$ of the crystal, we found that the band dispersion of the lowest conduction band and the highest valence band at wave-vector $(k_x,k_y)$ close to X valleys can be described by the following two-band Hamiltonian
\begin{widetext}
    \begin{align}
        H_\bk = \left[2(t_1+t_3){\rm cos}\left(\frac{a_yk_y}{2}\right){\rm cos}\left(\frac{a_xk_x}{2}\right)+t_2\right]\sigma_x + \left[2(t_3-t_1){\rm sin}\left(\frac{a_yk_y}{2}\right){\rm cos}\left(\frac{a_xk_x}{2}\right)\right]\sigma_y + \Delta \sigma_z \label{eq:2bnd_model},
    \end{align}
\end{widetext}
where $a_x$ and $a_y$ are the lattice constants along the zigzag and armchair directions, $t_i$ ($i=1,2,3$) represents the hopping parameters of the Se-$p_x$ orbitals to their nearest Sn-$p_x$ orbitals (see the sketch in Fig. S1) and $\Delta$ accounts for the energy difference between the energy of orbitals on each atomic site, and  $\sigma_i$ ($i=x,y,z$) indicates Pauli matrices acting on the two-band basis. Details of the derivation are provided in the section S1 of the SM. Fitting the parameters of this two-band model with the DFT band structure (see Fig. S2), we found that this minimal model can describe the band dispersion well in the vicinity of the X valleys. More importantly, from this model we can derive an analytical expression for the degree of circular dichroism at the X valleys
\begin{align}
    \eta_X =  \frac{2(t_1^2-t_3^2)a_xa_y{\rm sin} (\frac{1}{2}a_xk_{x0}){\rm cos} (\frac{1}{2}a_xk_{x0})}{(t_1+t_3)^2a_x^2{\rm sin}^2(\frac{1}{2}a_xk_{x0})+(t_1-t_3)^2a_y^2{\rm cos}^2(\frac{1}{2}a_xk_{x0})}, \label{eq:eta_tbm}
\end{align}
where $k_{x0}=\frac{2}{a_x}{\rm arccos}\left(-\frac{t_2}{2(t_1+t_3)}\right)$ is the position of the band edge predicted from Eq.~\eqref{eq:2bnd_model}. Evaluating $k_{x0}$ and the circular dichroism Eq.~\eqref{eq:eta_tbm} with the fitted $t_i$, we found $\eta_X\approx 30 \%$, agreeing well with DFT calculations (34\%) as shown in Fig. S3 of SM. This value sets an upper bound for the valley asymmetry, i.e. before considering any electron-phonon scattering, which would dynamically reduce this asymmetry. Furthermore, the model in Eq.~\eqref{eq:2bnd_model} and \eqref{eq:eta_tbm} can also be used to describe the circular dichroism in other valleytronic materials, for exemple graphene with broken inversion symmetry~\cite{YaoXiao2008}, where the three hopping parameters $t_i$ are equivalent due to symmetry, and Eq.~\eqref{eq:eta_tbm} gives a 100\% circular dichrosim, as expected. As a result, this analytical model also provides a symmetry insight to the valley asymmetry in SnSe.

\section*{Conclusions}
In conclusion, we have established quantum-geometric chiroptical selection rules as a mechanism for valley-selective light–matter coupling in a material class that lies outside the conventional hexagonal-valley paradigm. By combining time- and angle-resolved extreme-ultraviolet photoemission spectroscopy with time-dependent Boltzmann equation simulations, we directly imaged a helicity-dependent population imbalance between the X and X' valleys of SnSe, demonstrating that circularly polarized light can address valleys related by time-reversal symmetry even in a puckered, low-symmetry lattice that lacks the three-fold rotational symmetry generally invoked to explain near-unity circular dichroism in TMDCs. This shows that chiroptical valley selectivity is not a special feature of hexagonal monolayers, but a more general consequence of momentum-resolved Berry curvature wherever crystal symmetry allows it, extending the reach of circular-dichroism-based valleytronics to a broader family of van der Waals materials.

This chiroptical addressability adds a distinct and complementary control knob to the valleytronics of group-IV monochalcogenides. Where linearly polarized light selects between valleys along orthogonal high-symmetry directions (X/Y) via their different orbital character, circularly polarized light instead discriminates between the degenerate, time-reversal-partner X and X' valleys via their opposite Berry curvature. The two mechanisms are independent and, in principle, combinable, so that polarization state alone, i.e., linear angle and circular helicity, can be used to program which valley, and to what degree, is populated. This gives SnSe and related MX compounds access to a richer valley-manipulation phase space than materials relying on a single valley pair.

\section*{Acknowledgement}
We acknowledge Yann Mairesse for insightful discussions. We thank Nikita Fedorov, Romain Delos, Pierre Hericourt, Rodrigue Bouillaud, Laurent Merzeau, and Frank Blais for technical assistance.  We thank Baptiste Fabre for implementing and maintaining the data binning code. We acknowledge the financial support of the IdEx University of Bordeaux/Grand Research Program ``GPR LIGHT". This work is part of the ULTRAFAST and TORNADO projects of PEPR LUMA and was supported by the French National Research Agency, as a part of the France 2030 program, under grants ANR-23-EXLU-0002 and ANR-23-EXLU-0004. We acknowledge support from ERC Starting Grant ERC-2022-STG No.101076639, Quantum Matter Bordeaux, AAP CNRS Tremplin, and AAP SMR from Université de Bordeaux. S.F. acknowledges funding from the European Union’s Horizon Europe research and innovation programme under the Marie Skłodowska-Curie 2024 Postdoctoral Fellowship No 101198277 (TopQMat). Q.C. acknowledges funding from the TERAQUANTUM project of the Région Nouvelle-Aquitaine. U.D. acknowledges funding from the European Union’s Horizon Europe research and innovation programme under the Marie Skłodowska-Curie grant No HORIZON-MSCA-2023-DN-01 101169225 - SPARKLE. Funded by the European Union. Y.P and F.C acknowledge the funding by Deutsche Forschungsgemeinschaft (DFG) -- project number 443988403. The simulations are performed thanks to the computing time provided by the high-performance computer Lichtenberg at the NHR Centers NHR4CES at TU Darmstadt (Project p0021280). Views and opinions expressed are however those of the author(s) only and do not necessarily reflect those of the European Union. Neither the European Union nor the granting authority can be held responsible for them. 

\section*{Author contributions}
S.B. conceived the research project. S.F., U.D., Q.C. and S.B. performed the experiments. S.B. analyzed the experimental data. D.D. and S.P. participated in maintaining the laser system. Y.P. and F.C. developed the theoretical framework. Y.P. performed simulations and numerical implementations. S.B. and Y.P. wrote the first draft of the manuscript. Y.P., S.F., U.D., Q.C., D.D., F.C. and S.B. participated in commenting and revising the manuscript.

\section*{Data availability}
The data supporting the findings of this article will be made openly available on Zenodo upon publication.

\section*{Code Availability} 
Codes and programs for reproducing the theoretical calculations are available from the corresponding author upon reasonable request.

\end{document}